\documentclass[aps,rmp,longbibliography,reprint,nofootinbib,superscriptaddress]{revtex4-1}

\newcommand{\VERSION}{h} 

\newcommand{\referencelist}[1]{\bibliography{#1}}

\usepackage{graphicx,dcolumn,bm,amsmath,amssymb,euscript}
\usepackage[utf8]{inputenc}

\usepackage[colorlinks,citecolor=red,urlcolor=blue]{hyperref}

		\usepackage[update,prepend]{epstopdf}    

\newcommand{\Angstrom}{\text{\AA}}   

\begin{document}

\title[Coherence at APS]{Characterization of the X-ray Coherence Properties of an Undulator Beamline at the Advanced Photon Source}

\author{Guangxu Ju}
\author{Matthew J. Highland}
\affiliation{Materials Science Division, Argonne National Laboratory, Argonne, IL 60439}

\author{Carol Thompson}
\affiliation{Department of Physics, Northern Illinois University, DeKalb IL 60115}

\author{Jeffrey A. Eastman}
\author{Paul H. Fuoss}
\affiliation{Materials Science Division, Argonne National Laboratory, Argonne, IL 60439}

\author{Hua Zhou}
\affiliation{X-ray Science Division, Argonne National Laboratory, Argonne, IL 60439}

\author{Roger Dejus}
\affiliation{Accelerator Systems Division, Argonne National Laboratory, Argonne, IL 60439}

\author{G. Brian Stephenson}
\email[correspondence to: ]{stephenson@anl.gov}
\affiliation{Materials Science Division, Argonne National Laboratory, Argonne, IL 60439}

\date{revision \VERSION  : printed \today}

\begin{abstract}
In anticipation of the increased use of coherent x-ray methods and the need to upgrade beamlines to match improved source quality, we have characterized the coherence properties of the x-rays delivered by beamline 12ID-D at the Advanced Photon Source. 
We compare the measured x-ray divergence, beam size, brightness, and coherent flux at energies up to 26 keV to the calculated values from the undulator source, and evaluate the effects of beamline optics such as a mirror, monochromator, and compound refractive lenses. 
Diffraction patterns from slits as a function of slit width are analyzed using wave propagation theory to obtain the beam divergence and thus coherence length. 
Imaging of the source using a compound refractive lens was found to be the most accurate method for determining the vertical divergence.
While the brightness and coherent flux obtained without a monochromator (``pink beam'') agree well with those calculated for the source, those measured with the monochromator were a factor of 3 to 6 lower than the source, primarily because of vertical divergence introduced by the monochromator.
The methods we describe should be widely applicable for measuring the x-ray coherence properties of synchrotron beamlines.

\end{abstract}

\maketitle


\section{Introduction}

Coherent x-ray methods are providing revolutionary new capabilities for observing nanoscale dynamics and imaging atomic structure in materials. For example, x-ray photon correlation spectroscopy (XPCS) \cite{2009_stephenson_naturematerials,2014_Shpyrko_JSynchRad21_1057} has revealed the dynamics of atomic diffusion in crystals \cite{Lietner_NM_2009} and glasses \cite{ruta2014revealing} and atomic steps on electrode/electrolyte interfaces \cite{Karl_PCCP_2015}.
Likewise, coherent diffractive imaging methods \cite{2013_Abbey_JOM65_1183} are revealing strain distributions and dislocations by using Bragg scattering \cite{yau2017bragg} and surface steps by using crystal truncation rod scattering \cite{Zhu_APL_2015}.
Beamlines specifically designed for coherent x-ray methods are enhancing these studies \cite{chubar2011application,fluerasu2011analysis,rau2011coherent,2012_Winarski_JSynchRad19_1056}.

The Advanced Photon Source (APS) and other synchrotron facilities worldwide are being upgraded or built to provide greatly increased coherent x-ray flux, using a multi-bend achromat storage ring lattice \cite{2014_Hettel_JSR21_843}. As part of these efforts, new beamlines are being designed that will take advantage of the high coherent flux \cite{APSU_ConceptualDesign,ESRFU_ConceptualDesign}, and optics on existing beamlines are being enhanced to preserve the coherent flux from the source \cite{yabashi2014optics}. In anticipation of these developments, we are exploring coherent x-ray methods at APS beamline 12ID-D using the current source. We would like to deploy coherent x-ray methods at relatively high x-ray energies (e.g. $> 25$ keV) to enable {\em in situ} studies of materials processing. Like most beamlines at APS, 12ID-D was not originally designed for coherent x-ray methods. Thus we are interested in understanding which beamline optics will need to be improved as part of the APS Upgrade project.

In this paper we characterize the coherence properties of the x-rays delivered by this beamline at higher energies, and compare these to the calculated values from the source. We evaluate effects of beamline optics such as a high-heat-load mirror, monochromator, and compound refractive lenses. Previous studies have characterized the divergence (and thus the transverse coherence length) of the x-rays at synchrotron beamlines by measuring the apparent size of the source, using either a small aperture as a ``pinhole'' camera \cite{borland1989,millsNIM1990,elleaumeJSR1995,1996_Cai_RSI67_3368,sandy1999design} or focusing optics to produce an image \cite{cai1997beam}. Here we use both of these methods to determine divergences: measurements of vertical and horizontal diffraction patterns from slits, and measurements of the vertical focus size produced by a compound refractive lens. With the former method, we go beyond previous work by varying the slit width to quantify both the Fraunhofer (small-width) and Fresnel (large-width) limits and optimize the intermediate pinhole image case. With the latter method, we focus both the monochomatic beam and the pink beam (without a monochromator), to characterize the effect of the monochromator on the vertical divergence. Measured beam sizes and fluxes are combined with these divergences to obtain emittance, brightness, and coherent flux. All of these quantities are compared with values calculated for the undulator source, to understand the effect of beamline optics. The methods we describe to characterize coherence properties are of general applicability to synchrotron x-ray beamlines.

\section{Expressions for Coherence Properties}

The transverse coherence length $\xi$ is inversely related to the beam divergence (angular spread through a point) $r$, and is proportional to the wavelength $\lambda$.
The relationship between $\xi$ and $r$ includes a numerical factor that differs among references
\cite{1991_Sutton_Nature352_608,1997_Libbert_JSynchRad4_125,1997_Vlieg_JSynchRad4_210,2012_Jacques_PRB86_144117}, 
primarily because it matters whether $\xi$ and $r$ are considered to be root-mean-square values (e.g. the standard deviation $\sigma$ of a Gaussian distribution), or are considered to be full-width-at-half-maximum (FWHM) values.
Here we will use FWHM values, and the formula
\begin{equation}\label{eq:xi}
\xi = \frac{\lambda}{2 r}.
\end{equation}
The divergence $r$ as well as the overall beam size $s$ typically change as a function of distance from the source. For perfect optics that accept the full size of the beam, the conserved quantity is the product of divergence and beam size, known as emittance $\epsilon$.
\begin{equation}
    \epsilon \equiv r s
\end{equation}
In the simplest case where the optics deflect the beam only in the vertical or the horizontal, we expect the vertical and horizontal emittances $\epsilon_v \equiv r_v s_v$ and $\epsilon_h \equiv r_h s_h$ to be separately conserved.

Another quantity conserved by perfect optics is the spectral brightness $B$. Using FWHM values for the divergences and sizes entering into the emittances, the expression for $B$ is
\begin{equation}\label{eq:B}
B = \frac{F}{\epsilon_v \epsilon_h},
\end{equation}
where $F$ is the spectral flux.
The spectral flux can be obtained from the total flux $F_{tot}$ by accounting for the energy bandwidth used in the measurement,
\begin{equation}\label{eq:F}
F = \left ( \frac{0.1 \%}{\Delta \lambda / \lambda} \right ) F_{tot}, 
\end{equation}
where 0.1\% is the standard bandwidth for spectral flux or brightness, and $\Delta \lambda / \lambda$ is the experimental bandwidth for $F_{tot}$.
Imperfections in the optics can decrease the delivered brightness by reducing the spectral flux or increasing the emittances.

The spectral coherent flux $F_{coh}$ is proportional to brightness and wavelength squared,
\begin{equation}\label{eq:Fcoh}
F_{coh} = \left ( \frac{\lambda}{2} \right )^2 B 
= \left ( \frac{\xi_h}{s_h} \right )
\left ( \frac{\xi_v}{s_v} \right ) F. 
\end{equation}
One can see that the coherent fraction $F_{coh}/F$ is simply given by the ratios of the transverse coherence lengths $\xi$ to the beam sizes $s$ in the two directions.

\section{Measurements of beam divergence}

In the far field of the source, the beam divergence $r$ is a measure of the (apparent) angular size of the source.
Here we characterize the divergence of the x-ray beam delivered to the 12ID-D hutch by two methods:
measuring the diffraction pattern from slits as a function of slit size; and measuring the focal line width created by a compound refractive lens imaging the source.
The second method is found to be more sensitive \cite{cai1997beam}, and it is required to observe the effect of the monochromator on the vertical divergence. 

\begin{figure}[h]
\includegraphics[width=\columnwidth]{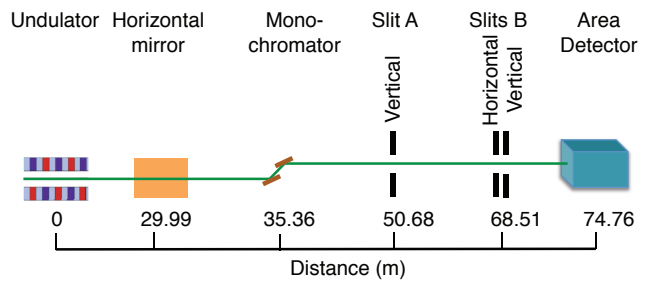}
\caption{A schematic of the beamline optics at 12ID-D and the experimental setup for measuring divergence using slits.}
\label{fig:schem_beamline_slits} 
\end{figure}

Figure \ref{fig:schem_beamline_slits} shows a schematic of the x-ray optics at APS beamline 12ID-D and the experimental setup for characterizing the vertical and horizontal divergences using diffraction from slits.
The x-ray source for the C and D stations is a 3.0 cm period undulator of 2.01 m effective length, located 1.25 m upstream of the nominal center of the straight section in the storage ring.
Distances from the source given in Fig.~\ref{fig:schem_beamline_slits} are from the center of this undulator.
The primary beamline optics consist of a horizontally deflecting mirror operating at a 4.2 mrad deflection angle,
as well as a Si (111) double-crystal cryogenically-cooled monochromator with a vertical diffraction plane \cite{Ramanathan_mono_1995}.
The mirror has active heating on its back side to counteract the beam heating and control its overall radius of curvature.
For these measurements, the mirror is positioned to use its Pt coating stripe,
which gives a critical energy for total reflection of $E_c = 40.5$ keV.
The beamline operates ``windowless,'' with a differential pump rather than a Be window separating the beamline vacuum from the storage ring vacuum.
Downstream, two unpolished 0.5-mm-thick Be windows are present in the setup, at 51 and 67 m from the source.
Other downstream flight path windows are made of materials with uniform electron density and smooth surfaces (e.g. Kapton film) to avoid any x-ray wavefront disturbance.

\subsection{Diffraction from Slits}

We analyzed diffraction patterns from slits to characterize the vertical and horizontal divergences of the x-ray beam.
Diffraction patterns were recorded on a high-resolution area detector consisting of a phosphor (Lu$_3$Al$_5$O$_{12}$:Ce) imaged onto a high-resolution scientific CMOS camera having 2560 $\times$2160 pixels of spacing 6.5~$\mu$m
(Andor Technologies Neo sCMOS, Model DC 152Q-COO-FI).
The optical magnification was 7.5, giving an effective x-ray pixel size at the phosphor of $p = 0.868$ $\mu$m.
This detector was located at the back of the 12ID-D hutch, at a distance of 74.76 m from the source.
Diffraction patterns were measured from slits at two different distances from the source and detector.
Slits B, located at the front of the 12ID-D hutch at $L_s = 68.51$~m from the source and $L_d = 6.25$~m upstream from the detector, were used to measure both vertical and horizontal diffraction patterns.
They were JJ X-ray AT-F7-AIR ESRF-style crossed slits with modified slit edges.
Tungsten carbide cylinders with 2.4 mm diameter were mounted along the edges of the slit blades to provide straight, polished edges.  
Slits A, located in the beamline at $L_s = 50.68$~m from the source, were also used to measure vertical diffraction patterns with a larger slit-to-detector distance, $L_d = 24.08$~m. These slits had flat tungsten knife edges.

\begin{figure}
\includegraphics[width=\columnwidth]{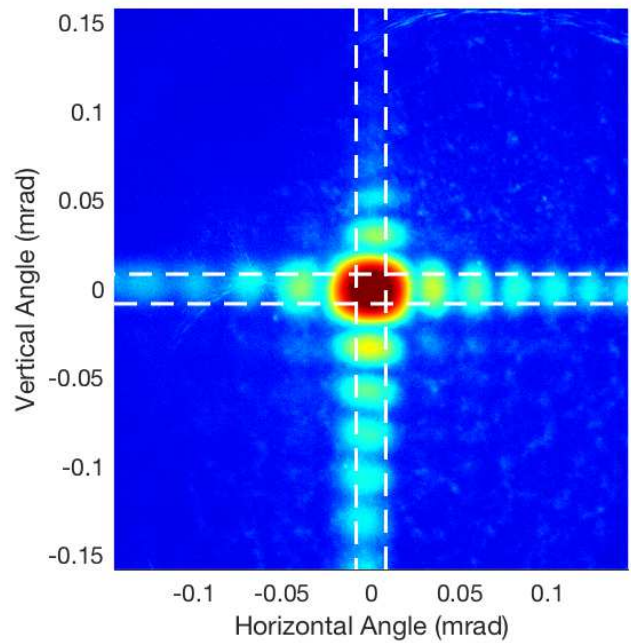}
\caption{Typical diffraction pattern from narrow crossed slits B at 22.1 keV.
Redder hues indicate higher intensities (log scale).
Regions integrated to obtain vertical and horizontal intensity profiles are shown by dashed lines.}
\label{fig:fraun2D} 
\end{figure}

\begin{figure}
\includegraphics[width=0.8\columnwidth]{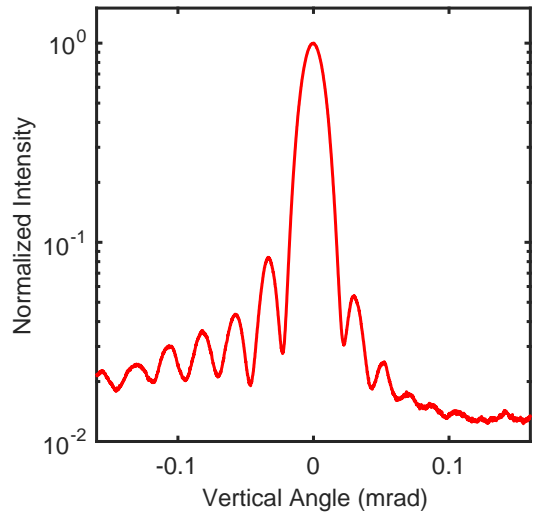}
\caption{Vertical intensity profile extracted from Fig. \ref{fig:fraun2D}.}
\label{fig:2015_1118_1_scan65_vertical} 
\end{figure}

\begin{figure}
\includegraphics[width=2in]{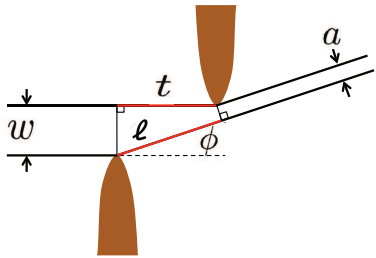}
\caption{A schematic of the geometry of the asymmetric slits used for diffraction measurements (with expanded vertical scale), showing the slit width $w$, slit offset $t$, diffraction angle $\phi$, the maximum path length difference $\Delta \equiv \ell - t$, and apparent slit width $a$.}
\label{fig:schem_asymslits} 
\end{figure}

Figure~\ref{fig:fraun2D} shows a typical diffraction pattern from slits B as a function of diffraction angle $\phi$. 
Here $\phi$ is obtained using an angular spacing per detector pixel of $p/L_d$.
The crossed slits produce a central peak surrounded by fringes in both directions.
Since rectangular slits produce diffraction patterns that are simply the product of those from linear slits in each direction, we can extract the separate patterns from the vertical and horizontal slits by integrating in the transverse direction.
Regions shown by dashed lines in Fig.~\ref{fig:fraun2D} were used to obtain separate vertical and horizontal diffraction patterns.
Figure~\ref{fig:2015_1118_1_scan65_vertical} shows the resulting vertical diffraction pattern.
The observed diffraction patterns show a high background level that is not present in the calculated patterns shown in Fig.~\ref{fig:wave_prop} below, which likely arises from light scattering within the detector phosphor. 
Nonetheless, the high spatial resolution of this detector system allows accurate determination of the fringe positions and the central peak width.
Analysis of the fringe spacing can be used to accurately determine the slit width $w$ in the Fraunhofer region, while analysis of the width of the central peak as a function of $w$ can be used to determine (or put an upper limit) on the divergence $r$ of the beam incident on the slits.

\subsubsection{Fraunhofer Diffraction from Asymmetric Slits}

Here we obtain the relation needed to extract the slit width from the fringe spacing in diffraction patterns from asymmetric slits.
In Figs.~\ref{fig:fraun2D} and \ref{fig:2015_1118_1_scan65_vertical}, it is apparent that the diffraction pattern is not symmetric about zero, and detailed examination shows that the fringe spacing varies with angle.
This asymmetric pattern arises because the opposite blades of the vertical and horizontal slits are offset from each other along the beam direction so that the blades do not collide when the aperture closes completely.
Figure~\ref{fig:schem_asymslits} shows the geometry of such asymmetric slit blades and the definitions of the slit width $w$, the offset $t$, the diffraction angle $\phi$, and other quantities used in the diffraction analysis below.
Slits A and B have offsets of $t = 30$ and $3.8$~mm, respectively. 

Such asymmetric slits have been used and analyzed in several previous x-ray studies
\cite{1987_Lang_JPhysD20_541,1997_Libbert_JSynchRad4_125,1997_Vlieg_JSynchRad4_210, 2002_Bolloch_JSynchRad9_258}, and their far-field (Fraunhofer) diffraction pattern has been calculated.
The diffraction intensity from an asymmetric linear slit in the Fraunhofer limit is given by \cite{2002_Bolloch_JSynchRad9_258}
\begin{equation}\label{eq:Fraun}
I_{\text{Fraun}}(\phi) \propto \left ( \frac{\sin (kw^{\prime}\phi/2)}{kw^{\prime}\phi/2} \right )^2,
\end{equation}
where  $k = 2 \pi / \lambda$ is the x-ray wavenumber. 
This result is similar to the standard formula for Fraunhofer diffraction from a symmetric linear slit, but with the slit width $w$ replaced by an effective slit width $w^{\prime}$. In the limit of small $\phi$, the effective width is given by \cite{2002_Bolloch_JSynchRad9_258}
\begin{equation}\label{eq:wp}
    w^{\prime} \approx w - t \phi / 2,
\end{equation}
which depends on the diffraction angle $\phi$.
This result can be simply understood by writing the Fraunhofer intensity as
\begin{equation}\label{eq:Fraun2}
I_{\text{Fraun}}(\phi) \propto \left ( \frac{\sin (k\Delta/2)}{k\Delta/2} \right )^2,
\end{equation}
where $\Delta$ is the path length difference between rays incident on each edge of the slit. 
From Fig. \ref{fig:schem_asymslits}, one can calculate $\Delta$ as
\begin{eqnarray}
    \Delta &\equiv& \ell - t \nonumber \\
    &=& w \sin \phi + t \cos \phi - t \nonumber \\
    &\approx& w \phi - t \phi^2 / 2.
\end{eqnarray}

From Eqs.~\ref{eq:Fraun} and \ref{eq:Fraun2}, one can see that the Fraunhofer diffraction pattern has a central peak at $\phi = 0$, with a FWHM in $\phi$ of 0.886$\lambda/w$, surrounded by fringes spaced in $\phi$ by $\lambda/w^{\prime}$.
For asymmetric slits (non-zero $t$), the spacing of the fringes varies with $\phi$, because $w^{\prime}$ varies with $\phi$. 
The positions of the minima that separate the fringes are given by $\phi_n = n \lambda/w^{\prime}$, where $n$ is a positive or negative integer. 
At these positions, the path length difference across the slit is a multiple of the wavelength, $\Delta = n \lambda$. 
The index $n$ is related to the angles $\phi_n$ of these minima by
\begin{equation}\label{eq:min_vs_phi}
n = (w \phi_n - t \phi_n^2/2) / \lambda,
\end{equation}
which contains a quadratic term proportional to $t$ as well as a linear term proportional to $w$.

It is compelling to attribute the fringe spacing
variation with $\phi$ to the change in the apparent slit width $a$ as a function of viewing angle, as shown in Fig.~\ref{fig:schem_asymslits}. 
In this case one would simply modify the standard Fraunhofer formula by replacing the slit width $w$ by its apparent width $a$ \cite{1997_Vlieg_JSynchRad4_210}. 
However, this produces a correction that is off by a factor of 2, since $a$ is not equal to $w^{\prime}$ of Eq.~\ref{eq:wp}, 
\begin{eqnarray}
a &=& w \cos \phi - t \sin \phi \nonumber \\
    &\approx& w - t \phi.
\end{eqnarray}
A fit to data using Eq.~(\ref{eq:Fraun}) with $a$ instead of $w^{\prime}$ and allowing $t$ to vary will give a value of $t$ that is a factor of 2 smaller than the actual offset.

\begin{figure}
\includegraphics[width=\columnwidth]{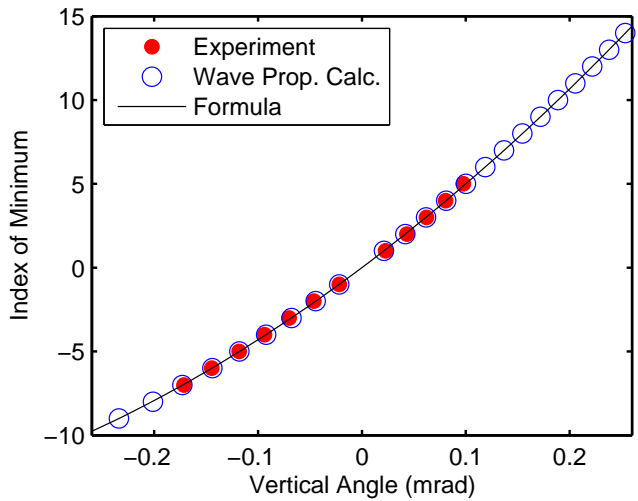}
\caption{Fitting of observed positions of minima in the diffraction pattern in Fig. \ref{fig:2015_1118_1_scan65_vertical} (red solid circles) to obtain slit B width $w = 2.61~\mu$m, using the formula given in Eq.~\ref{eq:min_vs_phi} (black curve) with a slit offset of $t = 3.8$~mm. Also shown (blue open circles) are the minima from a wave propagation theory calculation using these values of $w$ and $t$.}
\label{fig:fit_min2vangle} 
\end{figure}

We recorded sets of diffraction patterns at various slit width settings at x-ray photon energies of 12.0 and 22.1 keV $(\lambda = 1.033$ and 0.561 $\Angstrom)$, using the first and third harmonics of the undulator, respectively.
We determined the actual slit widths $w$ for the diffraction patterns that corresponded to the far-field limit $(w << (0.886 \lambda L_d)^{1/2})$ by extracting the positions of several minima $\phi_n$ at positive and negative $n$, and fitting to Eq.~\ref{eq:min_vs_phi}.
Figure~\ref{fig:fit_min2vangle} shows a typical fit, allowing the slit width $w$ and the zero value of the measured $\phi$ to vary.
The quadratic dependence of the index on angle given by the formula agrees well with the observed values.

The differences between the uncalibrated slit width positioner settings and the actual widths $w$ were found to be constant to within 1.5 or 0.5 $\mu$m FWHM for slits A or B, respectively, indicating that the main inaccuracy of the positioner setting was a simple offset.
For the diffraction patterns with larger slit widths not satisfying the far-field criterion, the width $w$ could not be determined as described above. 
Therefore we used a width obtained by correcting the positioner settings by the offset determined from the smaller slit widths.
This gave adequate accuracy for the larger slit widths.

\subsubsection{Wave Propagation Theory for Diffraction from Asymmetric Slits}

To analyze the full width of the central peak as a function of $w$ to determine the divergence $r$ of the incident beam, we require a theory that applies to the full range of slit widths spanning the far-field and near-field regions (Fraunhofer and Fresnel diffraction).
We can calculate such diffraction patterns by propagating the optical wave field from the source, through the slits, to the detector in the paraxial approximation using Fourier methods \cite{Fresnel}.  
Here we use a spherical wave from a point source, propagating a distance $L_s$ to the first slit blade, then a distance $t$ to the second slit blade, then a distance $L_d$ to the detector.
At each position along the beam direction, the wavefield can be expressed as an amplitude that varies in the transverse coordinate.
At each slit blade position, we calculate the incident wavefield amplitude propagated from the previous position, and then the region of the wavefield blocked by the blade is set to zero. 
Calculations are conveniently made in MATLAB on a discrete lattice of points transverse to the beam direction using fast Fourier transforms (FFTs) \cite{Fresnel}.
The amplitude propagated to the detector is then squared to give the diffracted intensity in spatial coordinates. 
These are converted to standard angular coordinates $\phi$ by dividing by $L_d$, to obtain $I_{\text{wp}}(\phi)$ for an ideal point source.

To include the effect of a finite source size, we convolute $I_{\text{wp}}$ with a finite angular resolution function $R$,
\begin{equation}
I_{\text{conv}}(\phi) = \int{I_{\text{wp}}(\phi^{\prime}) R(\phi - \phi^{\prime})} d\phi^{\prime},
\end{equation}
where here we approximate the resolution function by a Gaussian,
\begin{equation}
R(\Delta \phi) = \frac{1}{\sigma_R \sqrt{2\pi}} \exp \left ( \frac{-\Delta \phi ^2}{2 \sigma_R^2} \right ).
\end{equation}
In principle this angular resolution contains contributions from both the divergence of the beam incident on the slit due to the finite source size, and the exit angle resolution due to the finite spatial resolution of the detector.
However, the high spatial resolution of the detector system used corresponds to a negligible angular contribution ($p/L_d = 0.14~\mu$rad or $0.04~\mu$rad for Slits A or B, respectively).
Thus any measured broadening of the central peak above the ideal width of $I_{wp}$ provides a measure of the source size and thus the incident beam divergence $r$ at the slit position.
Since $r$ is defined as the FWHM of the angular distribution, it is related to $\sigma_R$ by
$r = 2 (2 \ln 2)^{1/2} \sigma_R$.

\begin{figure}
\includegraphics[width=0.8\columnwidth]{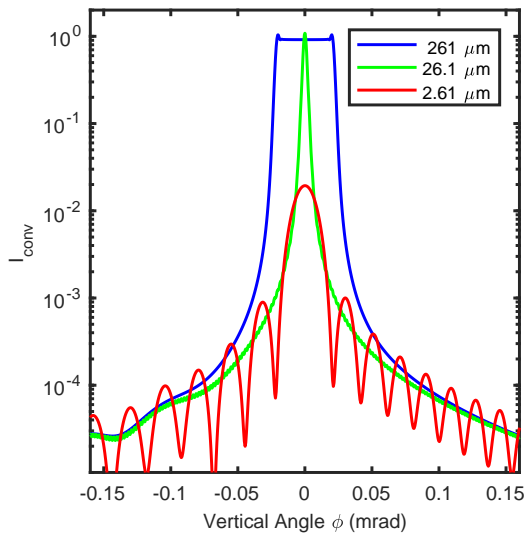}
\caption{Vertical intensity profiles $I_{\text{conv}}(\phi)$ calculated using wave propagation theory for slits B at 22.1 keV using the widths $w$ given in the legend, $t = 3.8$~mm, $r = 2$ $\mu rad$.}
\label{fig:wave_prop} 
\end{figure}

Figure \ref{fig:wave_prop} shows typical calculated diffraction patterns $I_{\text{conv}}(\phi)$ for slit B with various widths, using an incident beam divergence of $r = 2~\mu$rad
and energy 22.1 keV. The widths are chosen to span a range from the Fraunhofer to the Fresnel limit. 
In the Fraunhofer limit, we see a similar variation in fringe spacing with angle as in the experiments.
To check the consistency of the wave propagation analysis with the Fraunhofer formula for the minima, Eq.~\ref{eq:min_vs_phi}, we determined the positions of the fringe minima obtained from $I_{\text{conv}}$ calculated for the parameters corresponding to the data in Fig.~\ref{fig:fit_min2vangle}.
These are plotted as open circles in Fig.~\ref{fig:fit_min2vangle},
and they agree well with the measured data and the formula.

\begin{figure}
\includegraphics[width=\columnwidth]{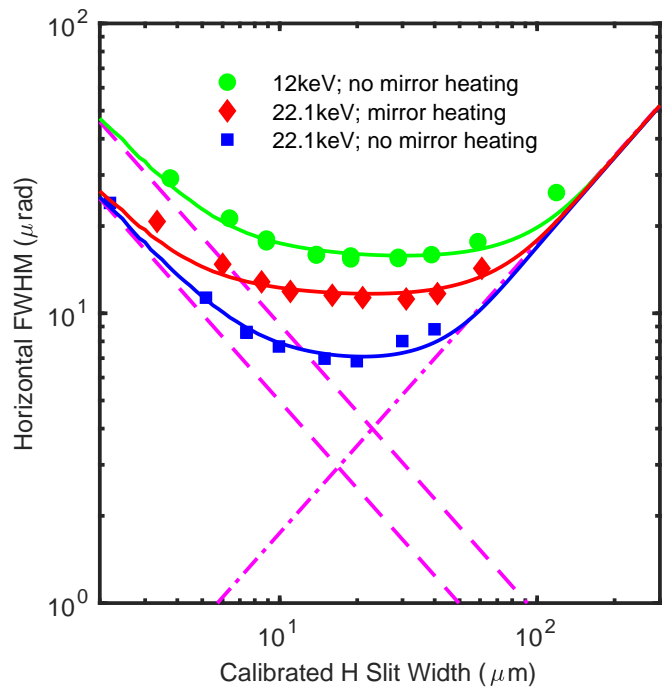}
\caption{Symbols: Measured horizontal full-width at half maximum of central peak as a function of calibrated slit width, at 12.0 and 22.1 keV, for slits B.
At 22.1 keV, data were measured for two values (0 and 10 W) of heating power on the beamline horizontal mirror.
Curves: calculated FWHM, for best-fit incident divergences given in Table~\ref{tab:hordiv}, using actual slit and source positions.
Dashed lines: diffraction limit for perfect resolution for each wavelength. Dash-dot line: slit width contribution to resolution.
}
\label{fig:FWHM_vs_Hslit} 
\end{figure}

\begin{figure}
\includegraphics[width=\columnwidth]{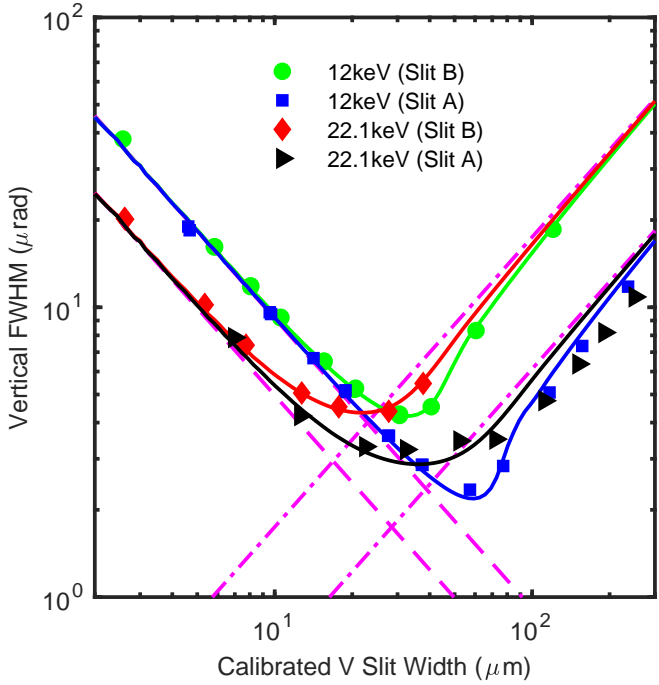}
\caption{Symbols: Measured vertical full-width at half maximum of central peak as a function of calibrated slit width, at 12.0 and 22.1 keV, for slits A and B.
Curves: calculated FWHM, for best-fit incident divergences given in Table~\ref{tab:vertdiv}, using actual slit and source positions.
Dashed lines: diffraction limits for perfect resolution for each wavelength. Dash-dot lines: slit width contributions to resolution for each slit.
}
\label{fig:FWHM_vs_Vslit} 
\end{figure}

To obtain the incident beam divergence at the slit position, we analyzed the FWHM of the central peak in the measured diffraction patterns as a function of the calibrated slit width. 
These values are shown as symbols in Figs.~\ref{fig:FWHM_vs_Hslit} and \ref{fig:FWHM_vs_Vslit} for the horizontal and vertical directions, respectively.
For the horizontal direction, Fig.~\ref{fig:FWHM_vs_Hslit}, patterns were measured using slit B at 12.0 and 22.1 keV; we also investigated the effect of the curvature of the beamline horizontally deflecting mirror, as controlled by heating its back side.
For the vertical direction, Fig.~\ref{fig:FWHM_vs_Vslit}, patterns were measured using both slits A and B at both 12.0 and 22.1 keV.
Mirror heating did not affect the vertical intensity profiles.

\begin{figure}
\includegraphics[width=\columnwidth]{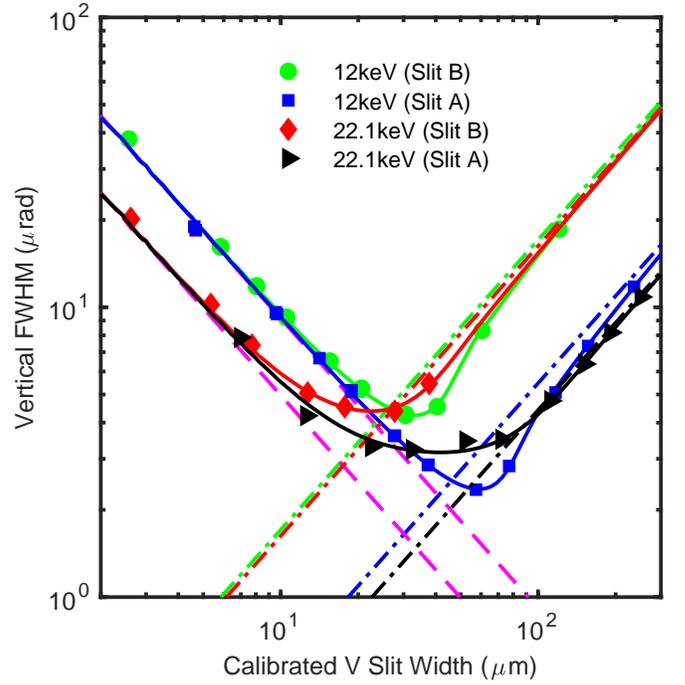}
\caption{Same data as in Fig.~\ref{fig:FWHM_vs_Vslit}, fit to calculated curves by varying incident divergences (results given in Table~\ref{tab:vertdiv}), using effective source positions of 100 and 400~m upstream of the detector at 12.0 and 22.1 keV, respectively.
Dashed lines: diffraction limits for perfect resolution for each wavelength. Dash-dot lines: slit width contributions to resolution for each slit and effective source position.
}
\label{fig:FWHM_vs_Vslit_2} 
\end{figure}

The dashed lines in Figs.~\ref{fig:FWHM_vs_Hslit} and \ref{fig:FWHM_vs_Vslit} show the diffraction limit $r_d = 0.886 \, \lambda /w$ for the far-field (Fraunhofer) region at small $w$.
The dash-dot lines show the $\phi$ angle corresponding to the projection of the slit width onto the detector, $r_w = w(L_s + L_d)/(L_s L_d)$, which gives the asymptotic value for the near-field (Fresnel) region at large $w$.
On a log scale, these asymptotes form limiting ``V" shapes for the divergence, with a minimum at $r_d = r_w$ given by
\begin{equation}\label{eq:rmin}
    r_{min} = \left [ \frac{0.886 \, \lambda (L_s + L_d)}{L_s L_d}\right ]^{1/2},
\end{equation}
at a slit width of 
\begin{equation}
    w_{min} = \left [ \frac{0.886 \, \lambda L_s L_d}{(L_s +  L_d)}\right ]^{1/2}.
\end{equation}
At this slit width, the detector is at the boundary between the near-field and far-field regions of the slit.
If the beam divergence $r$ incident on the slit exceeds $r_{min}$, its contribution will dominate over those of $r_d$ and $r_w$ for slit widths near $w_{min}$, and it can be accurately determined from the observed FWHM of the slit diffraction.
In this case, the central peak forms a ``pinhole camera'' image of the source.
The best resolution (minimum $r_{min}$) for fixed beamline length $L_s + L_d$ occurs with the aperture half way to the source, $L_s = L_d$, giving $r_{min} = [3.54 \, \lambda / (L_s + L_d)]^{1/2}$, $w_{min} = [0.222 \, \lambda (L_s + L_d)]^{1/2}$.

The solid curves in Figs.~\ref{fig:FWHM_vs_Hslit} and \ref{fig:FWHM_vs_Vslit} show the FWHM of the central peak of $I_{\text{conv}}$ calculated from wave propagation theory convoluted with various divergences $r$. 
The value of $r$ for each curve has been adjusted to best fit the corresponding data points.
For the horizontal direction, Fig.~\ref{fig:FWHM_vs_Hslit}, the incident beam divergences are larger than $r_{min}$ in all three cases, so that the central peak FWHM at the minimum is determined by a pinhole camera image of the source.
For the vertical direction, Fig.~\ref{fig:FWHM_vs_Vslit}, the incident beam divergences are not sufficiently larger than $r_{min}$ to clearly distinguish their contributions.

\begin{table}
\caption{ \label{tab:hordiv} Horizontal divergences $r_h$ obtained from fits in Fig.~\ref{fig:FWHM_vs_Hslit} (all at position of slits B, with monochromator). Also shown are calculated coherence lengths $\xi_{h}$.} 
\begin{ruledtabular}
\begin{tabular}{ c | c | c | c | c}   
 x-ray energy  & mirror heating &  $r_h$ & limit $r_{min}$ & $\xi_h$\\
 (keV) & (W) & ($\mu$rad) & ($\mu$rad) & ($\mu$m)\\
\hline
 12.0 & 0 & 14.9 & 4.0 & 3.5\\
 22.1 & 0 & 6.3  & 2.9 & 4.5\\
 22.1 & 10 & 11.0 & 2.9 & 2.6\\
\end{tabular}
\end{ruledtabular}
\end{table}

\begin{table}
\caption{ \label{tab:vertdiv} Vertical divergences $r_v$ at position of slits A or B obtained from fits in Figs.~\ref{fig:FWHM_vs_Vslit} and \ref{fig:FWHM_vs_Vslit_2} (all with monochromator). Also shown are calculated coherence lengths $\xi_v$.} 
\begin{ruledtabular}
\begin{tabular}{c | c | c | c | c | c | c}   
 & x-ray energy  & method & $L^{\text{eff}}_s$ &  $r_v$ & limit $r_{min}$ & $\xi_v$   \\
 & (keV)    &   & (m) & ($\mu$rad) & ($\mu$rad) & ($\mu$m) \\
\hline
Fig.~\ref{fig:FWHM_vs_Vslit} & 12.0 & slits B & 68.51 & (2.8) & 4.0 & (18.5)\\
& 22.1 & slits B & 68.51 & (3.4) & 2.9 & (8.2)\\
& 12.0 & slits A & 50.68 & 1.3 & 2.4 & 39.7\\
& 22.1 & slits A & 50.68 & 2.3 & 1.7 & 12.2\\
\hline
Fig.~\ref{fig:FWHM_vs_Vslit_2} & 12.0 & slits B & 93.75 & (2.9) & 4.0 & (17.8)\\
& 22.1 & slits B & 393.75 & (3.5) & 2.8 & (8.0)\\
& 12.0 & slits A & 75.92 & 1.6 & 2.2 & 32.3\\
& 22.1 & slits A & 375.92 & 2.7 & 1.5 & 10.4\\
\end{tabular}
\end{ruledtabular}
\end{table}

In Fig.~\ref{fig:FWHM_vs_Vslit}, one can see that the measured data fall below the best fit curves and the limiting asymptotes $r_w$ at all large slit widths, especially for slit A.
One explanation for this effect is that wavefront distortions introduced by the monochromator increase the effective distance to the source, so that the projection of the slit width onto the detector is smaller than expected.
Figure~\ref{fig:FWHM_vs_Vslit_2} shows calculated curves using effective source distances $L^{\text{eff}}_s$ corresponding to source positions $L^{\text{eff}}_s + L_d = 100$ and 400~m at 12.0 and 22.1 keV, respectively. 
These values were varied along with $r$ to give the best fits to the data.
Allowing the effective source positions to vary avoids skewing the $r_v$ values to attempt to fit the deviations at large $w$ apparent in Fig.~\ref{fig:FWHM_vs_Vslit}. 
Thus we expect that the $r_v$ values obtained from the fits in Fig.~\ref{fig:FWHM_vs_Vslit_2} are most reliable.

Tables~\ref{tab:hordiv} and \ref{tab:vertdiv} give the best fit values for the horizontal and vertical beam divergences $r_h$ and $r_v$, respectively, obtained from the fits.
These are compared with the calculated minimum divergences $r_{min}$ at the intersections of the asymptotes.
All of the fit values of $r_h$ exceed the corresponding $r_{min}$ values, and thus should be reasonably accurate.
Because the values of $r_{min}$ for slit B are larger than those for slit A, we expect the values of $r_v$ from slit B to be less accurate, so we list them in parentheses.
In both tables, we also show the transverse coherence lengths calculated from the divergences using Eq.~(\ref{eq:xi}).

\subsection{Focal Line Width} 

\begin{figure}[h]
\includegraphics[width=\columnwidth]{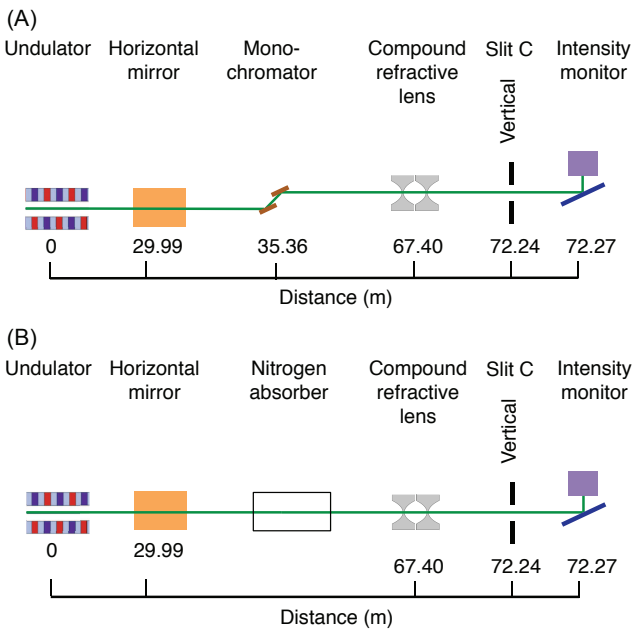}
\caption{A schematic of the beamline optics at 12ID-D and the setups for measuring vertical divergence by imaging the source using a compound refractive lens. (A) with a monochromator; (B) without a monochromator (pink beam). }
\label{fig:schem_beamline_crl} 
\end{figure}

Because of the small vertical source size, the divergence of the incident beam in the vertical is too small to definitively measure using slit diffraction, even using slits A located in the beamline a large distance from the detector.
We therefore investigated an alternative method, using a focusing optic to image the apparent source size.
Figure~\ref{fig:schem_beamline_crl} shows schematics of the setups used to measure vertical focal line widths, both with the cryogenically cooled Si (111) monochromator in place, and with the monochromator removed (``pink beam'').
In both cases, a vertically focusing compound refractive lens (CRL) consisting of a set of 42 individual double-concave Be lenses with 200 $\mu$m tip radii (obtained from RXOPTICS Refractive X-RAY Optics, Monschau, Germany) was installed at a distance $L_s = 67.40$~m from the source.
A $100~\mu$m vertical aperture was placed in front of the CRL.
The undulator gap was tuned so that its 3rd harmonic peaked at 25.75 keV.
For this x-ray energy, an image of the vertical source size was formed at a distance $L_d = 4.84$~m downstream of the CRL.
We measured the vertical focal line width by scanning a small slit C through the focus.
Intensities were monitored upstream of the slit using an ion chamber and downstream of the slit using a glass cover slip scattering into a PIN diode. 

\begin{figure}
\includegraphics[width=\columnwidth]{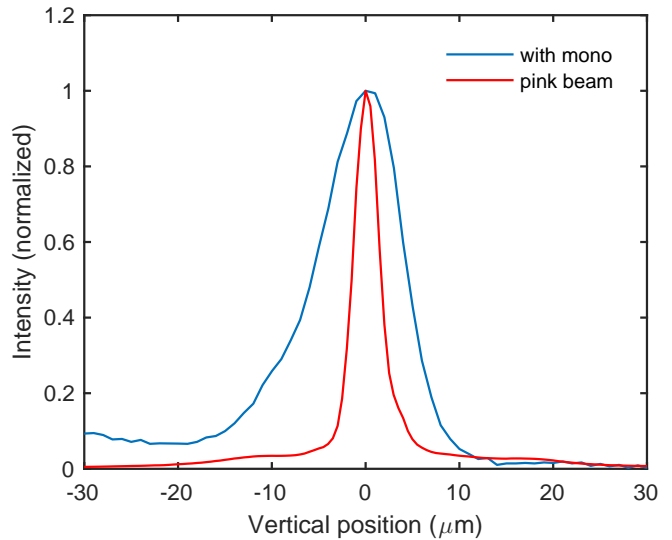}
\caption{Comparison of vertical focus profiles with monochromatic and pink beam at 25.75 keV. Observed widths were 10.6 and 3.2 $\mu$m FWHM, respectively.}
\label{fig:mono_pink_Vfocus} 
\end{figure}

For the monochromatic case, the combination of monochromator and horizontal mirror effectively removed harmonics at energies other than 25.75 keV.
For the pink beam case, an 8 m length of flight path upstream of the CRL was filled with N$_2$ gas at 1 atm, to absorb most of the lower (first and second) harmonics of the undulator spectrum.
Because of the strong $E^2$ dependence of the CRL focal length on photon energy, only the 25.75 keV third harmonic of the undulator contributed significantly to the sharp focus.
The contribution to the focal line width of the energy bandwidth of the third harmonic can be estimated as $s_v^{CRL} \Delta \lambda / \lambda$, where $s_v^{CRL}$ is the size of the aperture at the CRL.
Using the radial width of a Bragg peak from an analyzer crystal, we measured the bandwidth of the undulator third harmonic to be $\Delta \lambda / \lambda = 8.5 \times 10^{-3}$. For $s_v^{CRL} = 100~\mu$m, this gives a contribution of only $0.85~\mu$m to the focal line width.
The contribution from diffraction at the aperture, $0.886 \, \lambda L_d / s_v^{CRL}$, was $2.1~\mu$m.

\begin{table}
\caption{ \label{tab:CRLv} Vertical divergences $r_v$ and coherence lengths $\xi_{v}$ obtained from focal line width measurements using the compound refractive lens at $L_s = 67.40$~m and 25.75 keV, with and without the monochromator.} 
\begin{ruledtabular}
\begin{tabular}{c | c | c | c | c }   
mono & x-ray energy  & method &  $r_v$ &  $\xi_{v}$  \\
 & (keV)    &   & ($\mu$rad) & ($\mu$m) \\
\hline
Yes& 25.75 & CRL & 2.1 & 11.3\\
No & 25.75 & CRL & 0.4 & 57.3\\
\end{tabular}
\end{ruledtabular}
\end{table}

Figure \ref{fig:mono_pink_Vfocus} shows measured focal line profiles at 25.75 keV with the monochromator in place, and with it removed.
The divergence of the beam incident on the CRL is given by
\begin{equation}
    r = \frac{s_{src}}{L_s} = \frac{s_{foc}}{L_d},
\end{equation}
where $s_{src}$ and $s_{foc}$ are the sizes (FWHM) of the apparent source and determined focus.
The measured focal line widths with the monochromator in place and removed were 10.6 and 3.2~$\mu$m FWHM, respectively.
We deconvoluted the contributions from the measurement slit C size, aperture diffraction, and, in the case of pink beam, bandwidth, to obtain
$s_{foc} = 9.9$ and 2.0~$\mu$m, which correspond to divergences of 2.06 and 0.42 $\mu$rad, respectively, at the CRL.
As summarized in Table \ref{tab:CRLv}, the much smaller value obtained with pink beam indicates that the monochromator introduces significant vertical divergence.

The pink beam divergence of 0.42 $\mu$rad measured with the CRL is significantly smaller than the limiting values $r_{min}$ available with slit diffraction.
Thus, aberrations in the CRL which contribute to the measured divergence are evidently sufficiently small that this method can measure much smaller beam divergences than possible with slits.

\begin{figure}
\includegraphics[width=3in]{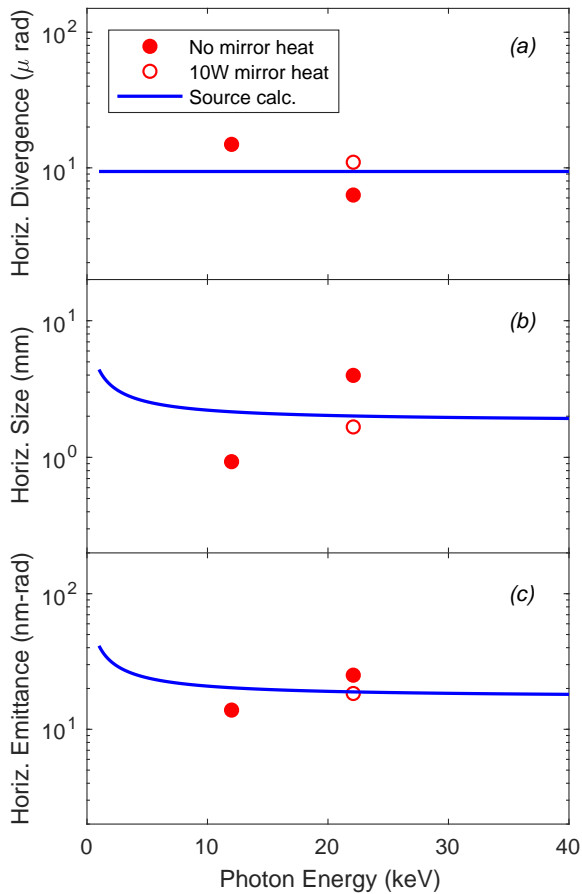}
\caption{Comparison of measured 12ID-D horizontal divergence, size, and emittance (symbols) to the ideal values from the source (curves) as a function of photon energy, for a source distance of $L_s = 68.51$~m.}
\label{fig:horizontal_div_emit} 
\end{figure}

\begin{figure}
\includegraphics[width=3in]{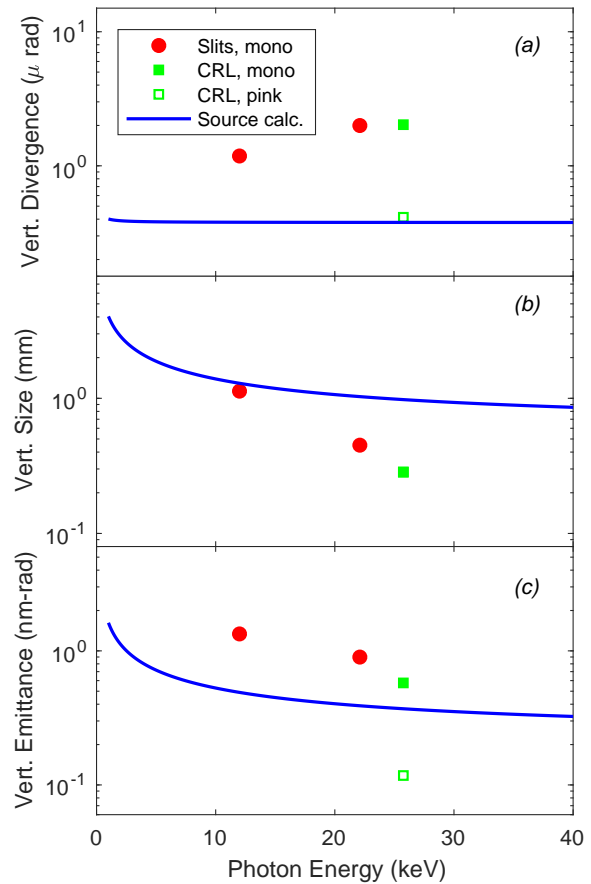}
\caption{Comparison of measured 12ID-D vertical divergence, size, and emittance (symbols) to the ideal values from the source (curves) as a function of photon energy, for a source distance of $L_s = 68.51$~m.}
\label{fig:vertical_div_emit} 
\end{figure}

\subsection{Summary of Divergence and Emittance Measurements}

To obtain the emittances from the divergences measured using slits, we also measured the beam sizes by scanning slit B, set to a small aperture, across the beam and monitoring the transmitted intensity.
(For the focal width measurements, we scanned slit C with the CRL removed.)
Here we summarize the measured divergences, beam sizes, and emittances, and compare them to ideal calculated values for the undulator source, as a function of photon energy.

In the far field of the source, the ideal divergence of the x-ray beam is determined by the size of the source, and the ideal size of the x-ray beam is determined by the divergence of the source \cite{borland1989}.
The r.m.s. (Gaussian sigma) electron beam source sizes $\sigma_e$ and divergences $\sigma_e^{\prime}$ for APS sector 12 with the 2.51 nm-rad lattice are
$\sigma_{eh} = 274$ $\mu$m, 
$\sigma_{ev} = 11.0$ $\mu$m, 
$\sigma_{eh}^{\prime} = 11.3$ $\mu$rad,
and
$\sigma_{ev}^{\prime} = 3.6$ $\mu$rad,
for the horizontal and vertical, respectively \cite{SourcePointData_APSwebpage}.
The contribution to the source divergences from the intrinsic photon divergence
\cite{Orangexraybook_sec2-1}
for an undulator of length 
$\ell_u = 2.01$ m is
$\sigma_r^{\prime} = (\lambda/\ell_u)^{1/2} = 7.2$ $\mu$rad (r.m.s) at 12 keV
and 5.4 $\mu$rad (r.m.s) at 22.1 keV.
The total source divergence for each direction is the sum in quadrature
of the electron beam and intrinsic photon terms,
$\sigma^{\prime} = (\sigma_e^{\prime 2} + \sigma_r^{\prime 2})^{1/2}$.
The intrinsic photon contribution to the source size is negligible
in our case, so that $\sigma = \sigma_e$.
In the far field at a distance $L_s$ from the source,
the ideal FWHM beam size
$s_0 = 2 (2 \ln 2)^{1/2} L_s \sigma^{\prime}$,
and the ideal FWHM incident beam divergence through a point 
is
$r_0 = 2 (2 \ln 2)^{1/2} \sigma / L_s$.
The ideal transverse emittance is given by
$\epsilon_0 = r_0 s_0 = (8 \ln 2) \sigma^{\prime} \sigma$.

Figure \ref{fig:horizontal_div_emit} summarizes the far-field divergences, sizes, and emittances in the horizontal direction at the position of slits B.
Because the beamline mirror can focus or defocus the beam in the horizontal, depending upon the x-ray heat load on its front surface and the active heating applied to its back, the measured divergence and beam size can be above or below the calculated source values that do not account for optics.
We see that when the undulator first harmonic is tuned to 12.0 keV (gap = 20.3 mm), giving a total power of 431 W and central power density of 10.6 kW/mrad$^2$, the horizontal divergence measured is larger than that calculated from the source, and the measured beam size is smaller, indicating a focusing effect of the beamline mirror when no heating is applied.
In contrast, when the undulator third harmonic is tuned to 22.1 keV (gap = 13.8 mm), giving a total power of 1807 W and central power density of 44.4 kW/mrad$^2$, the measured divergence is smaller and the measured beam size is larger than calculated, indicating a defocusing effect with no applied heating.
Applying a power of 10 W to the back of the mirror under these conditions slightly overcompensates the effect of beam heating.

The source emittance will be conserved by ideal focusing optics.
Indeed, the measured horizontal emittances are in good agreement with the source calculations for all three cases, indicating the mirror focusing is almost ideal.

Figure \ref{fig:vertical_div_emit} summarizes the far-field divergences, sizes, and emittances in the vertical direction.
For 12.0 and 22.1 keV, we show the divergence values obtained from slits A (the fit using an effective source distance), scaled by 50.68/68.51 to correspond to the position of slits B. 
Likewise the divergences and beam sizes obtained from the CRL line focus measurements are appropriately scaled to correspond to the position of slits B.
One can see that for monochromatic beam, the vertical divergences all exceed the calculated source value by a factor of 3 to 5, while with no monochromator (pink beam), the divergence is in excellent agreement with the calculation. 
This indicates that the monochromator is the primary source of increased vertical divergence.
The larger divergence at 22.1 and 25.75 keV relative to 12.0 keV could be related to the higher thermal load on the monochromator at these energies.

While the vertical beam size matches the calculation at 12.0 keV, it becomes successively smaller at higher energies. One possible contribution to this effect would be an overall curvature of the monochromator crystals, which can act to focus or defocus the beam \cite{AntimonovNIM2016}. The observed reduction in beam size at 25.75 keV would require a change in angle of ~20 $\mu$rad across the vertical beam profile. Another contribution could be aperturing due to misalignment of the angular bandpasses of crystals of two curvatures. As the Darwin width of the Si (111) reflection decreases from 23 $\mu$rad at 12.0 keV to 10 $\mu$rad at 25.75 keV \cite{Stepanov_website}, and the footprint of the beam on the crystal surfaces increases from 4.0 to 6.6 mm, an overall curvature difference of ~20 $\mu$rad over 5 mm would act as an effective aperture on the transmitted beam at higher energies.

The overall size of the pink beam from an undulator is much larger than that of the monochromatic beam at the on-axis peak of the undulator spectrum, because the radiation distribution from the undulator extends out to larger angles with an energy that decreases with increasing angle \cite{Orangexraybook_sec2-1}.
We did not independently measure the beam size of the 25.75 keV component of the pink beam, but simply used the size determined with the monochromator in place to calculate the emittance for the pink beam. 
This gives an emittance that is smaller than the calculated source value, consistent with an aperturing effect of the monochromator.

\section{Brightness and Coherent Flux}

We measured the x-ray flux using an air ion chamber at a storage ring current of 100 mA.
The total flux $F_{tot}$ in photons per second incident on an ion chamber is proportional to the ion chamber current $I_{ic}$ when the bias voltage is in the ``plateau'' region \cite{cherry2003physics}. 
The two can be related by
\begin{equation}
F_{tot} = I_{ic} / Q_{ph},
\end{equation}
where $Q_{ph}$ is the average charge created per photon.
This can be calculated from
\begin{equation}
Q_{ph} = e \left [ 1 - \exp \left ( \frac{-\ell_{ic}}{\ell_{pa}(E_{ph})} \right ) \right ] \left ( \frac{E_{ph}}{E_{ion}} \right ),
\end{equation}
where $e = 1.6 \times 10^{-19}$~C is the electronic charge, the second factor gives the fraction of photons absorbed in the collection length $\ell_{ic} = 6$~cm of the chamber, and the third factor gives the number of ions created by a photo-absorption event.
We use values of the photo-absorption length $\ell_{pa}$ calculated from the elemental photo-absorption cross-sections and gas densities for air at 1 atm (N$_{1.56}$ O$_{0.42}$ Ar$_{0.01}$)
\cite{1993_Henke_ANDT54_181,Henke_LBLURL}
and a value of $E_{ion} = 34.4$~eV \cite{Orangeraybook_sec4-5}.  
The values obtained for $Q_{ph}$ are
1.14, 0.30, and $0.22 \times 10^{-18}$~C at
12.0, 22.1, and 25.75 keV, respectively.
At 22.1 and 25.75 keV, we were able to reach the plateau of constant ion chamber current vs. voltage at $\sim$2300 V across a 12~mm gap.
At 12.0 keV, we were not able to reach the plateau, and we report values for the highest voltage of 3000 V.
We have found that these values correspond well with fluxes obtained using a He ion chamber operating in its plateau region.

\begin{figure}
\includegraphics[width=\columnwidth]{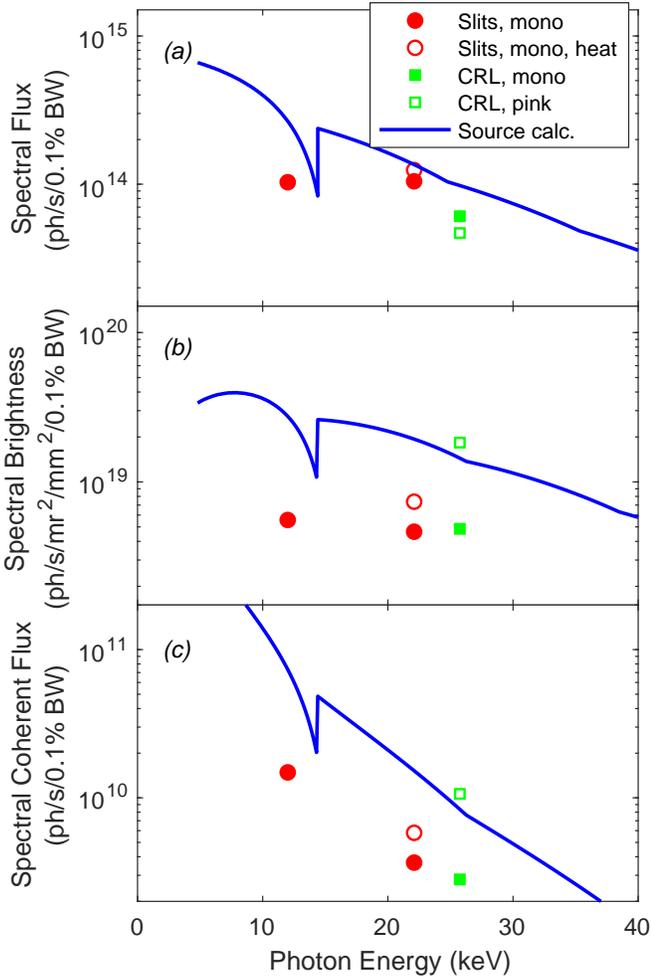}
\caption{Comparison of measured 12ID-D flux, brightness, and coherent flux (symbols) to the ideal values from the source (curves) as a function of x-ray energy for 100 mA ring current.}
\label{fig:bright_flux_coh} 
\end{figure}

Figure \ref{fig:bright_flux_coh} gives the measured and calculated spectral flux $F$, brightness $B$, and coherent flux $F_{coh}$.
Measured total fluxes have been converted to spectral flux using Eq.~(\ref{eq:F}) with a calculated bandwidth of $\Delta \lambda /\lambda = 0.131 \times 10^{-3}$
for the Si (111) monochromatic cases \cite{Stepanov_website}, or the measured value $\Delta \lambda /\lambda = 8.5 \times 10^{-3}$ for the pink beam case.
For monochromatic beam, we measured the total flux in the full beam size, and calculated $B$ and $F_{coh}$ using Eqs.~(\ref{eq:B}) and (\ref{eq:Fcoh}).
The horizontal divergence was not measured at 25.75 keV, but was estimated to be 22 $\mu$rad from the average horizontal emittance measured at 22.1 keV divided by the measured horizontal beam size of 1.0 mm at 25.75 keV.
For pink beam, we measured the on-axis total flux in an area $0.12 \times 0.10$~ mm (HxV) at $L_s = 68.5$~m to be $1.77 \times 10^{13}$ photons per second.
The value of $F$ plotted in Fig.~\ref{fig:bright_flux_coh}(a) was estimated by scaling to the measured monochromatic beam area at 25.75 keV.
Reasonable agreement between the spectral flux values for the pink and monochromatic cases indicates that this estimate is valid.
Pink beam brightness was calculated by dividing the spectral flux by this area, the measured pink vertical divergence, and the estimated monochromatic horizontal divergence.
Thus uncertainty in the overall size of the pink beam does not affect the calculated $B$ or $F_{coh}$ values.

The undulator performance was calculated using standard APS simulation codes.
The undulator parameters at 12ID-D were
3.0 cm period, 2.01 m effective length (effective $N = 67$),
and calculations were performed for a ring current of 100 mA. 
Values presented here are from the codes TC \cite{del2011xop} and TCAP
which account for the imperfections of a typical APS undulator.
We also ran calculations at specific energies
with code UR \cite{dejus1994program} using the measured magnetic field imperfections
in the specific 12ID-D undulator APS30\#5S;
the values obtained agree with those presented
within 2\% for brightness and 8\% for flux. 
The relationships given in Eqs.~(\ref{eq:B}-\ref{eq:Fcoh}) above between the brightness, flux, and beam parameters are obeyed within 2\% by the values presented.
These calculated values are significantly below those given by the analytical equations for an undulator with no imperfections \cite{Orangexraybook_sec2-1}, especially at higher x-ray energies.

The measured spectral fluxes are similar to or lower than the source calculations by up to a factor of 2.6.
For monochromatic beam, the measured brightness and coherent flux are lower than the calculations by a factor of 3 to 6, while for pink beam they agree well.
The lower spectral flux but almost equal brightness for pink beam, relative to the calculated values, is consistent with an aperturing effect on the monochromatic beam, since we used the monochromatic beam size in estimating the pink beam spectral flux.

\section{Summary and Conclusions}

In using diffraction from slits to characterize beam divergence, it was convenient to obtain accurate values for the slit width $w$ as a function of the positioner setting by measuring the spacing of the fringes in the Fraunhofer diffraction regime. 
We observed a variation in the fringe spacing as a function of diffraction angle that arises because of the non-zero offset $t$ of the asymmetric slits (Fig.~\ref{fig:schem_asymslits}). The relationship derived in Eq.~(\ref{eq:min_vs_phi}) agrees with experimental values and wave propagation theory (Fig.~\ref{fig:fit_min2vangle}). 

Using the Fraunhofer fringes to calibrate the slit widths, we obtained values of the FWHM of the central diffraction peak as a function of slit width (Figs.~\ref{fig:FWHM_vs_Hslit},~\ref{fig:FWHM_vs_Vslit}, and \ref{fig:FWHM_vs_Vslit_2}) that span from the Fraunhofer (small $w$) limit to the Fresnel (large $w$) limit. 
At intermediate values of $w$, the observed FWHM can be dominated by a pinhole camera image of the source size $s_{src}$, if the divergence $r = s_{src} / L_s$ from the source size is larger than the minimum observable divergence $r_{min}$ given in Eq.~(\ref{eq:rmin}).
The measured FWHM vs. slit width values were well fit by curves calculated from wave propagation theory, especially if the effective source distance was allowed to vary for the vertical case (Fig.~\ref{fig:FWHM_vs_Vslit_2}).
We were able to obtain accurate values for the horizontal divergence, since in this case $r$ is significantly larger than $r_{min}$.

More accurate values for the relatively small vertical divergence were obtained using an alternative method, imaging of the source with a CRL. 
After deconvoluting effects from aperture diffraction, detector resolution, and finite bandwidth, we obtained a divergence as small as 0.42~$\mu$rad at $L_s = 68.51$~m for pink beam at 25.75 keV with the monochromator removed, Fig.~\ref{fig:vertical_div_emit}(a), in excellent agreement with that calculated from the source.
Imaging the source with a focusing optic such as a CRL, rather than pinhole imaging with a slit, is required to measure the vertical divergence of current synchrotron sources such as APS. For future multi-bend achromat sources such as the APS Upgrade, this method will also be needed to determine the horizontal divergence, when the horizontal source size will be similar to the current vertical source size \cite{2014_Hettel_JSR21_843}.

Our measurements allow evaluation of the effects of various beamline optics on the coherence properties of the x-ray beam.
The horizontally-deflecting high-heat-load mirror, which was in place for all measurements, performed well to preserve the horizontal emittance of the source, as shown in Fig.~\ref{fig:horizontal_div_emit}(c), although it focused or defocused the beam depending upon the x-ray heat load on its front surface and the applied heating to its back surface.
This horizontally deflecting mirror did not affect the vertical divergence of the beam.
The requirements for the figure of such a mirror will be much more exacting after the APS Upgrade, with the much smaller horizontal source size. 

The good agreement of the measured pink beam vertical divergence and the source calculation indicates that the focusing CRL preserved the vertical emittance. Thus, CRLs will be good options for focusing coherent beams at higher energies such as 26 keV.

When the monochromator was in place, we observed vertical divergences that were factors of 3 to 5 higher than the calculated source values. 
While the vertical beam size was smaller than the calculated values at higher energies, the net effect was to produce a higher emittance at all energies investigated.
The total flux also tended to be lower than calculated.
These led to degradation of the brightness and coherent flux by factors of 3 to 6 below the calculated values.
Furthermore, fits of the slit diffraction central peak FWHM as a function of slit width (Fig.~\ref{fig:FWHM_vs_Vslit_2}) in the Fresnel region indicated an apparent source distance larger than the actual distance. 
All of these effects are consistent with aberrations in the monochromator \cite{AntimonovNIM2016} arising from thermal or mounting strains of the crystals.
Further work will be needed to understand and potentially improve the monochromator performance to deliver the full brightness and coherent flux.
The current monochromator is a typical APS cryogenic design \cite{Ramanathan_mono_1995}; other monochromator designs have been optimized for coherence preservation \cite{2012_Winarski_JSynchRad19_1056}.
Because the vertical divergence of the current APS source is similar to what it will be after the APS Upgrade, achieving coherent flux values approaching the ideal with a vertically diffracting monochromator and the current source will demonstrate the typical improvements needed to take full advantage of the revolutionary multi-bend achromat upgrade of the accelerator.

The brightness measured with pink beam provides a quantitiative validation of the undulator modeling codes, indicating that the very high brightness values at high energy calculated for APS Upgrade conditions will indeed be reached.
Furthermore, the current availability of high coherent flux at high energy using pink beam (e.g. $9 \times 10^{10}$ photons per second at 25.75 keV in a 0.85\% bandwidth) will allow us to begin exploring high energy coherent x-ray methods in experiments (such as XPCS from surface dynamics scattering near the specular direction) for which a wide bandwidth can be used. 
We can then look forward to more widely applying these techniques in the future using high-flux, high-energy monochromatic coherent beams  from the new coherent x-ray sources.

\begin{acknowledgments}
We would like to thank several APS staff for their expert assistance: Soenke Seifert and Chuck Kurtz for providing descriptions of the current optics at 12ID-D, Russell Woods for providing the area detector, and Xianbo Shi for the reference to code for wave propagation calculations.
This research was supported by the U.S. Department of Energy (DOE), Office of Science, Office of Basic Energy Sciences, Division of Materials Science and Engineering, and used resources of the Advanced Photon Source, a DOE Office of Science User Facility operated by Argonne National Laboratory.
\end{acknowledgments}



\referencelist{2018_12ID_coherence_char_h}  



\end{document}